\documentclass{llncs}
\usepackage{graphicx}
\usepackage{subfigure}
\usepackage{url}
\usepackage{common}

\begin{document}
\title{Aggregating Content and Network Information \\to Curate Twitter User Lists}
\titlerunning{TBD}
\authorrunning{Greene et al.}

\author{Derek Greene\inst{1} \and Gavin Sheridan\inst{2} \and Barry Smyth\inst{1} \and P\'{a}draig Cunningham\inst{1}}

\institute{School of Computer Science \& Informatics, University College Dublin 
\and Storyful, Dublin, Ireland}  


\maketitle

\begin{abstract}

Twitter introduced \emph{user lists} in late 2009, allowing users to be grouped according to meaningful topics or themes. Lists have since been adopted by media outlets as a means of organising content around news stories. Thus the curation of these lists is important -- they should contain the key information gatekeepers and present a balanced perspective on a story. Here we address this list curation process from a recommender systems perspective. We propose a variety of criteria for generating user list recommendations, based on content analysis, network analysis, and the ``crowdsourcing'' of existing user lists. We demonstrate that these types of criteria are often only successful for datasets with certain characteristics. To resolve this issue, we propose the aggregation of these different ``views'' of a news story on Twitter to produce more  accurate user recommendations to support the curation process.   

\end{abstract}

\section{Introduction}
Recently there has been a significant shift online towards the task of content curation and distillation, moving away from the traditional activity of content generation alone\footnote{\url{http://rww.to/y6TKoY}}. Notably, media outlets can now break or cover stories as they evolve by leveraging the content produced by users of social media sites (\eg videos, photographs, tweets). 
However, significant issues arise when trying to (a) identify content around a breaking news story in a timely manner, (b) monitor the proliferation of content on a certain news event over a period of time, and (c) ensure that this content is reliable and accurate. Storyful\footnote{\url{http://www.storyful.com}} is a social media news agency established in 2010 with the aim of filtering news, or newsworthy content, from the vast quantities of noisy data on social networks such as Twitter and YouTube. To this end, Storyful invests considerable time into the manual curation of content on these networks. In some cases this involves identifying key ``gatekeepers'' who are prolific in their ability to locate, monitor, and filter news from eyewitnesses. 

Twitter users can organise the users they follow into Twitter \emph{lists}. Storyful maintains lists of users relevant to a given news story, as a means of monitoring breaking news related to that story. Often these stories generate community-decided hashtags (\eg \emph{\#occupywallstreet}). But even with small news events, using such hashtags to track the evolution of a story becomes difficult. Spambots quickly intervene, while users with no proximity (in space, time or expertise) to the news story itself drown out other voices.
Manual curation via user lists is one way to overcome this problem, but this process is time-consuming, and risks incomplete coverage of all aspects of a news story. 

\begin{figure}[!t]
\centering
\includegraphics[width=0.97\linewidth]{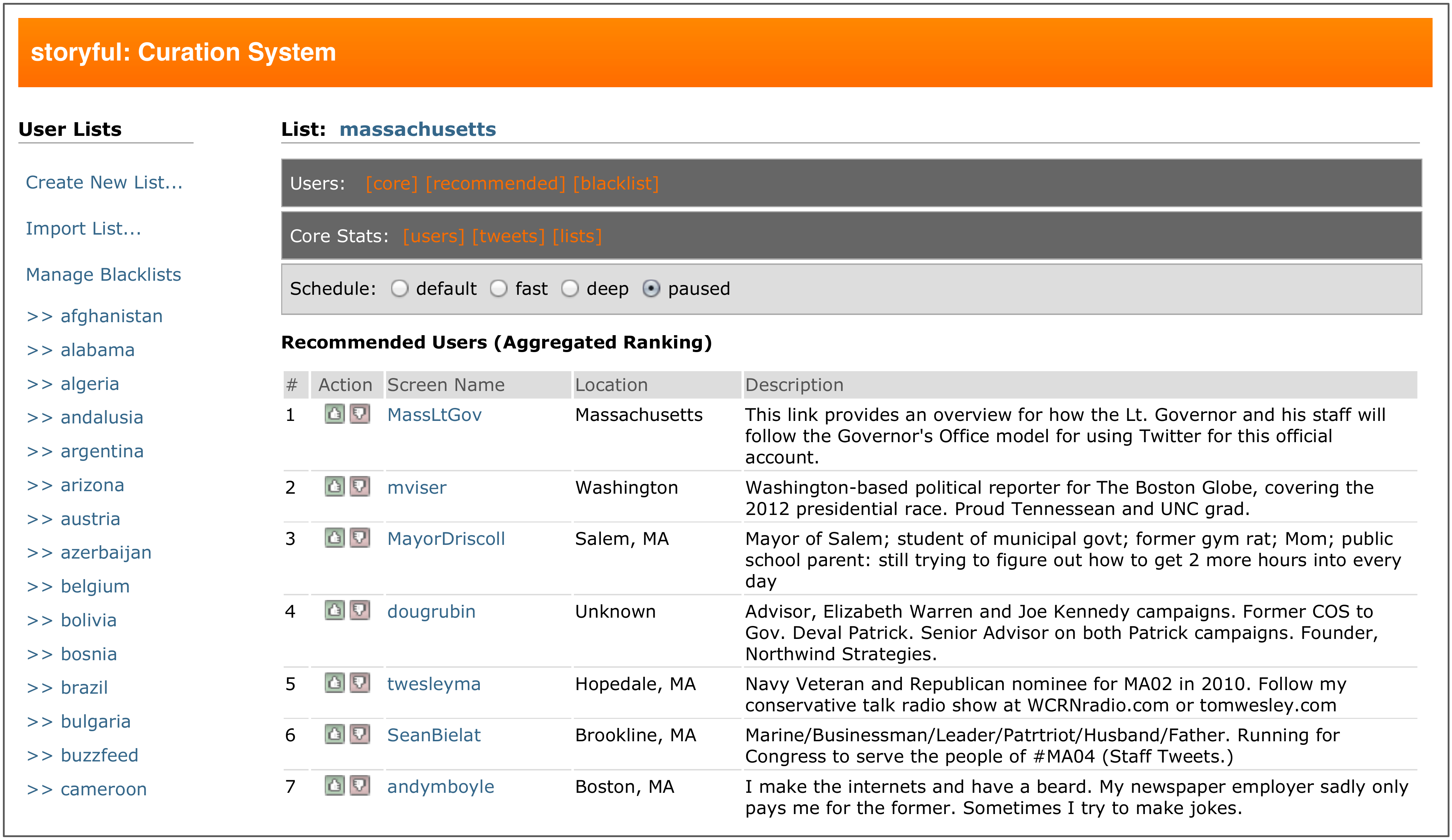}
\caption{Screenshot of the curation system, showing the presentation of candidate users recommended for addition to an existing user list covering the Republican nomination for the state of Massachusetts in March 2012.}
\label{fig:system}
\end{figure}

In this paper, we address Twitter list curation from a recommender systems perspective. The input to the recommendation process is an embryonic seed list, containing a small number of users that have been deemed to be  authoritative on the subject matter of the list by one or more journalists. While these \emph{seed users} may tweet on a variety of topics, the list defines the context in which recommendations should be made. For instance, when building a list about a political issue, if a journalist in the seed set tweets about politics and sport, then the recommendations should be concerned with politics rather than sport. To support the list curation tasks performed by Storyful, we have developed and deployed a system for exploring the Twitter network and recommending the important users that form the ``community'' around a news story (see \reffig{fig:system}).

In \refsec{sec:criteria} we propose a variety of network- and content-based criteria that are used to help to produce an expanded user list, which provides more comprehensive coverage of a news story. In \refsec{sec:eval} we evaluate these criteria on ten new datasets pertaining to Twitter discussions around the Republican Party nomination for the United States 2012 presidential election, where ground truth annotations have been provided by our partners at Storyful. Using a novel \emph{cohesion analysis} procedure, we demonstrate that, depending upon the Twitter data available, criteria based solely on either network or content analysis will not perform consistently well. Therefore, in \refsec{sec:eval2} we describe techniques for aggregating these criteria, and demonstrate that this aggregation process out-performs individual criteria across a range of datasets.

\section{Related Work}
\label{sec:related}

Many researchers have become interested in exploring content and network structure within Twitter, given the potential for the microblogging platform to facilitate the rapid spread of information. Kwak \etal \cite{kwak10twitter} studied a sample of 41.7 million users and 106 million tweets, investigating aspects such as: identifying influential users, information diffusion, and trending topics.  
Shamma \etal \cite{shamma09debates} performed an analysis on microblogging activity during the 2008 US presidential debates. The authors demonstrated that frequent terms reflected the topics being discussed, but the use of informal vocabulary complicated topic identification.

The specific problem of identifying influential and authoritative users on Twitter has been examined by a number of users. Kwak \etal \cite{kwak10twitter} performed initial work on ranking the importance of users, using both PageRank on the network of followers, and by counting the number of retweets achieved by each user.
Cha \etal \cite{cha10influence} attempted to capture different network perspectives by examining user follower in-degree, retweets, and mentions. Surprisingly, the authors found that follower information alone provided little evidence of authority, while the latter measures provided a better assessment of the level of engagement between users and their audience. Weng \etal \cite{weng10twitterrank} proposed a topic-specific adaptation of PageRank for ranking Twitter users according to their authority in a given area.

Researchers have also considered the related problem of producing personal recommendations for finding additional users to follow on Twitter, either by following network links or by performing textual analysis of tweet content. Hannon \etal \cite{hannon10twitter} proposed a set of techniques for producing personal recommendations on users to follow, based on the similarity of the aggregated tweets or ``profiles'' of users that are connected to the ego in the Twitter social graph. Such techniques have primarily relied on a single view of the network to produce suggestions. However, we can view the same Twitter network from a range of different perspectives.
For instance, Conover \etal \cite{conover11polar} performed an analysis of Twitter data based on references to other Twitter screen names in a tweet, while researchers have also looked at the diffusion of content via \emph{retweets} to uncover the spread of memes and opinions on Twitter \cite{conover11polar,rat11truthy}. The idea is that both \emph{mentions} and \emph{retweets} provide us with some insight of the differing interactions between microblogging users. 

In the context of enterprise social media, experiments performed by Daly \etal \cite{daly10network} suggested that, rather than automatically-generating social recommendations based on a single perspective, multiple recommender algorithms applied to different views may be preferable for supporting user selections.
\section{List Curation Criteria} 
\label{sec:criteria}
\hyphenation{attribut-able}

In this section we describe a variety of criteria used to generate recommendations based on Twitter data. Given a breaking or evolving news story, we begin with a small set of annotated seed users that has been manually identified by a curator. Our task then becomes to identify additional Twitter users relevant to the news story, in the form of a ranked list of recommended users provided to the curator.  

Each recommendation criterion involves representing the data in a sparse matrix representation, where users are represented by sparse \emph{profile vectors}.  To generate recommendations, we construct the mean or \emph{centroid} profile vector for the set of annotated training users, and then rank the test users according to the  cosine similarity between their profile vector and the centroid.

\subsection{Content-Based Criteria}

The popularity of these techniques may be partly attributable to the ability to apply existing techniques from text mining and information retrieval research to tweet content \cite{hannon10twitter}, and partly due to the ready availability of streaming tweet data (relative to other network-based Twitter data).

\begin{description}
\item[Tweet profiles.] Following the technique proposed by Hannon \etal \cite{hannon10twitter}, we construct a tweet profile vector for each user, consisting of the aggregation of a certain number of their most recent tweets. This results in a sparse term-user profile matrix. For the evaluations conducted in this paper, we consider up to the 50, 100 and 200 most recently-posted tweets for each user. As terms, we extract all unique words, hashtags, and user name mentions, while URLs are removed.
\item[List names.] Each Twitter user list has a human-readable \emph{name}, designated by the list creator, which usually indicates the topic to which the users in the list pertain (\eg ``Idaho Politics'', ``Machine Learning Researchers''). From this we derive a content-based criterion, where each user is represented by a term vector constructed from the aggregation of the tokens in names of the Twitter user lists to which they have been assigned. 
\item[List descriptions.] Twitter user lists can also have an optional \emph{description}, which often provides a more verbose definition of the type of user contained in the list. This allows us to derive another content-based criterion, where each user is represented by a term vector constructed from the aggregation of the \emph{descriptions} of the Twitter user lists to which they belong. 
\item[List merged.] Finally, to provide term vectors with a larger feature set, we can represent each user by a  vector constructed from the aggregation of the both \emph{names} {\bf{and}} \emph{descriptions} of the Twitter user lists to which they have been assigned.
\end{description}
Note that, for the content-based views, we apply standard log-based TF-IDF normalisation prior to generating recommendations.

\subsection{Network-Based Criteria}

We now describe a number of criteria based on network and graph views of Twitter. A motivating factor for these criteria is the use of co-citation information in bibliometrics research, which has been shown to often be more effective in revealing the true associations between papers than citations alone \cite{white81cocite}.

\begin{figure}[!t]
\centering
\includegraphics[width=0.73\linewidth]{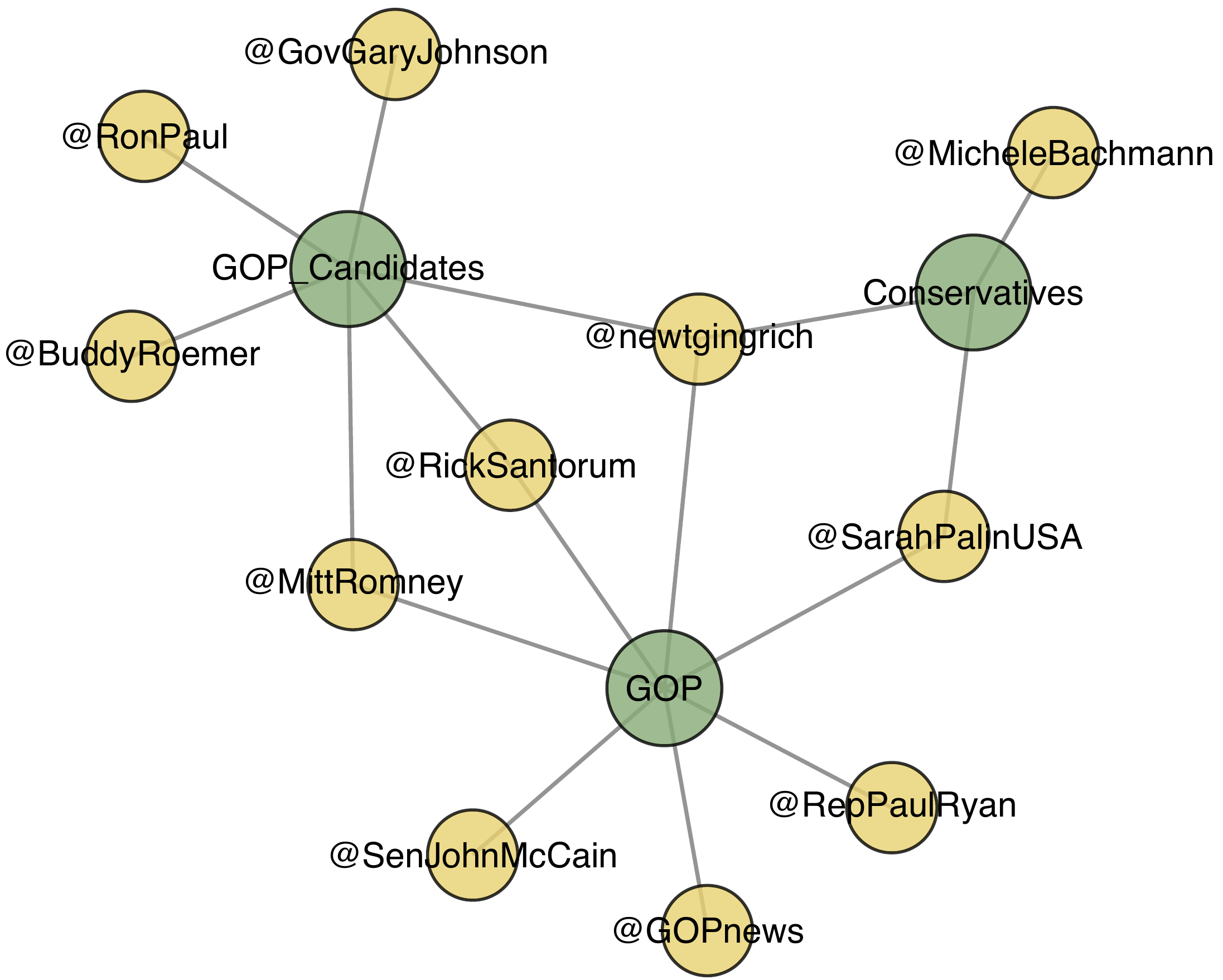}
\caption{Example of a bipartite user-list graph, containing eleven users assigned to three different user lists.}
\label{fig:list}
\end{figure}

\begin{description}
\item[Followed-by profiles.] The follower graph is an unweighted directed graph, where an edge exists from one node to another if one user follows another user. Reciprocal links exist where a pair of users follow one another. From this graph, for each user $u_{i}$ we can construct a sparse binary \emph{followed-by profile} vector $\vec{v}$, where an entry $v_{j}=1$ if user $u_{i}$ is followed by another user $u_{j}$, or $v_{j}=0$ otherwise. A pair of users are deemed to be similar if their vectors have a high cosine similarity -- \ie they are ``co-followed'' by the same users.\\

\item[Retweeted-by profiles.] The retweet graph is a weighted directed graph, where an edge exists from one node to another if one user retweets another user. From this graph, for each user $u_{i}$ we can construct a sparse real-valued \emph{retweeted-by profile} vector $\vec{v}$, where an entry $v_{j}$ indicates the number of times tweets posted by $u_{i}$ were retweeted by another user $u_{j}$. A pair of users are deemed to be similar if their vectors have a high cosine similarity -- \ie their tweets are frequently ``co-retweeted'' by the same users.\\

\item[Mentioned-by profiles.] The mention graph is a weighted directed graph, where an edge exists from one node to another if one user mentions another user. From this graph, for each user $u_{i}$ we can create a sparse real-valued \emph{mentioned-by profile} vector $\vec{v}$, where an entry $v_{j}$ indicates the number of times the user $u_{i}$ was mentioned in the tweets posted by another user $u_{j}$. A pair of users are deemed to be similar if their vectors have a high cosine similarity -- \ie they are ``co-mentioned'' by the same users. \\

\item[Co-listed information.]
Our primary motivation in this paper is to identify user list members relevant to a given news story. It may often be the case that other news organisations and private individuals will also be simultaneously curating user lists on the same topic in the wider Twittersphere. Ideally we would like to be able to ``crowdsource'' these efforts to support list curation. Based on existing Twitter user list memberships, we can construct a bipartite list-user graph, where an edge between a list and a user node indicates that the list contains the specified user. As an example,  \reffig{fig:list} shows a simple bipartite graph representing three user lists. The users \emph{@RickSantorum} and \emph{@MittRomney} are \emph{co-listed} twice, as both users are members of the lists \emph{GOP\_Candidates} and \emph{GOP}. If this co-listing is replicated across the wider Twitter network, this may be indicative of an affinity between the pair of users. Using a list-user matrix representation, we can compute the cosine similarity of users with one another, indicating the similarity of their list memberships profiles. Users who are more frequently co-listed will be deemed to be more similar. 
\end{description}

\section{Evaluation}
\label{sec:eval}

\subsection{Datasets}
\label{sec:data}

For evaluation purposes, we constructed a collection of Twitter datasets focused on news surrounding the Republican nomination for the United States presidential election of 2012. Specifically, these datasets focus on ten states where the votes were held on ``Super Tuesday'' (March 6, 2012). For each state, our partners at Storyful manually curated a set of between 20 and 97 seed users. For each seed user, we gathered a maximum of approximately 300 tweets, friends, followers, and user list memberships using the Twitter API. For any user list that we encountered, we also retrieved its associated name and description, if available. These limits reflect a quantify of data that might realistically be retrieved when monitoring multiple news stories in real-time, taking into account the comparatively strict query rate limits imposed by the Twitter API. 

For each dataset, we also generated an expanded set of non-seed users. These sets were created as follows: we constructed the follower graph for the seed users, ranked the non-seed users in the graph based on their in-degree, and selected up to 1,000 of these with the highest in-degree such that their in-degree was $\geq 2$. The rationale here is that these are prominent candidate members for the list in the Twitter neighbourhood of the story. Data was retrieved for these users using the same limits as used for the seed set.

This yielded ten datasets for evaluating the proposed curation criteria, containing on average $\approx 831$ total users, of which $\approx 5\%$ are annotated as seed users. In total 1,618,383 tweets from 8,305 unique users were collected. Details of these datasets are listed in \reftab{tab:data}. These datasets are made available online in pre-processed form\footnote{\url{http://mlg.ucd.ie/curation}}.

\begin{table*}[!t]
\caption{Summary of Twitter datasets used in our evaluations, including the number of annotated seed users and the total number of users. The four rightmost columns indicate the mean number of tweets, friends, followers, and list memberships per user in the complete dataset.}
\centering
\begin{tabular}{|l|cc|cccc|}\hline
\bf Dataset & \bf Seed Users & \bf Total Users & \bf Tweets & \bf Friends & \bf Followers & \bf Listed \\ \hline
\it Alaska & 41 & 948 & 185 & 208 & 269 & 89 \\ 
\it Georgia & 34 & 966 & 211 & 235 & 295 & 126 \\ 
\it Idaho & 20 & 743 & 186 & 264 & 273 & 47 \\ 
\it Massachusetts\;\;\;\; & 24 & 821 & 209 & 244 & 293 & 122 \\ 
\it North Dakota & 26 & 363 & 203 & 147 & 192 & 93 \\ 
\it Ohio & 97 & 1051 & 178 & 171 & 207 & 115 \\ 
\it Oklahoma & 32 & 693 & 205 & 178 & 211 & 109 \\ 
\it Tennessee & 48 & 979 & 199 & 170 & 204 & 112 \\ 
\it Vermont & 36 & 864 & 182 & 169 & 190 & 66 \\ 
\it Virginia & 46 & 877 & 200 & 160 & 199 & 115 \\ \hline
\end{tabular}
\label{tab:data}
\end{table*}

\subsection{Experimental Setup}
\label{sec:setup}

To compare the individual criteria introduced in \refsec{sec:criteria}, we perform multiple runs of k-fold cross-validation on each of the ten datasets, using the annotated seed users as a ground truth. For each fold, we hold out a proportion of the seed set for use as test data, and used the remaining seed users as training data in conjunction with the centroid-based recommender.  The goal of the recommendation task becomes that of distinguishing the users in the test set from the remaining non-seed users in the complete dataset, which constitute false positives.

We rank the non-training users using each criterion, and compute \emph{precision} and \emph{recall} scores relative to the test data for the top $k \in [10,50]$ recommendations. This process is repeated for 250 randomised runs, from which mean precision and recall scores are calculated. Note that, due to the differing number of seed users in each dataset, the number of folds for a given dataset is selected from $\in [2,5]$ so as to ensure there is at least ten users in each test set. 

\begin{figure}[!t]
\centering
\includegraphics[width=0.66\linewidth]{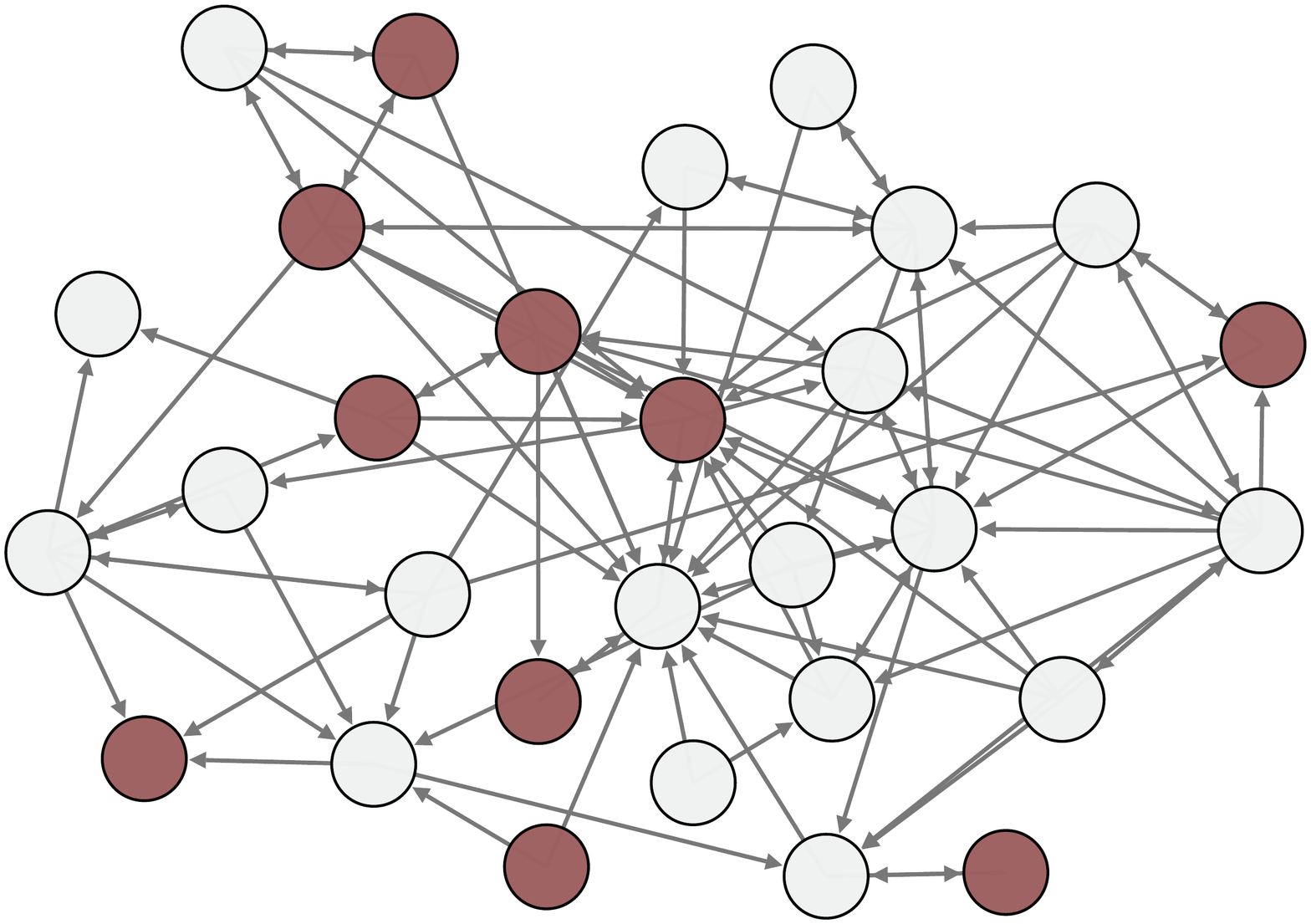}
\caption{Example of a single instance of 3-fold cross-validation on the subgraph induced by the seed set on the Georgia mentions network. The ten red (dark) nodes denote the users which have been held out as the test set. Note that the additional 934 false positives in the Georgia data are not shown here.}
\label{fig:fold}
\end{figure}

\reffig{fig:fold} shows an example of a single instance of 3-fold cross-validation on the subgraph induced by the seed set for the Georgia mentions network. The goal here is to identify the ten red (dark) nodes representing the set of users which have been held out as test data, based on the training set of blue (light) users, from among the larger set of 966 users in the complete dataset.

\subsection{Comparison of Criteria}
\label{sec:eval1}

\begin{figure}[!t]
\centering
\includegraphics[width=0.72\linewidth]{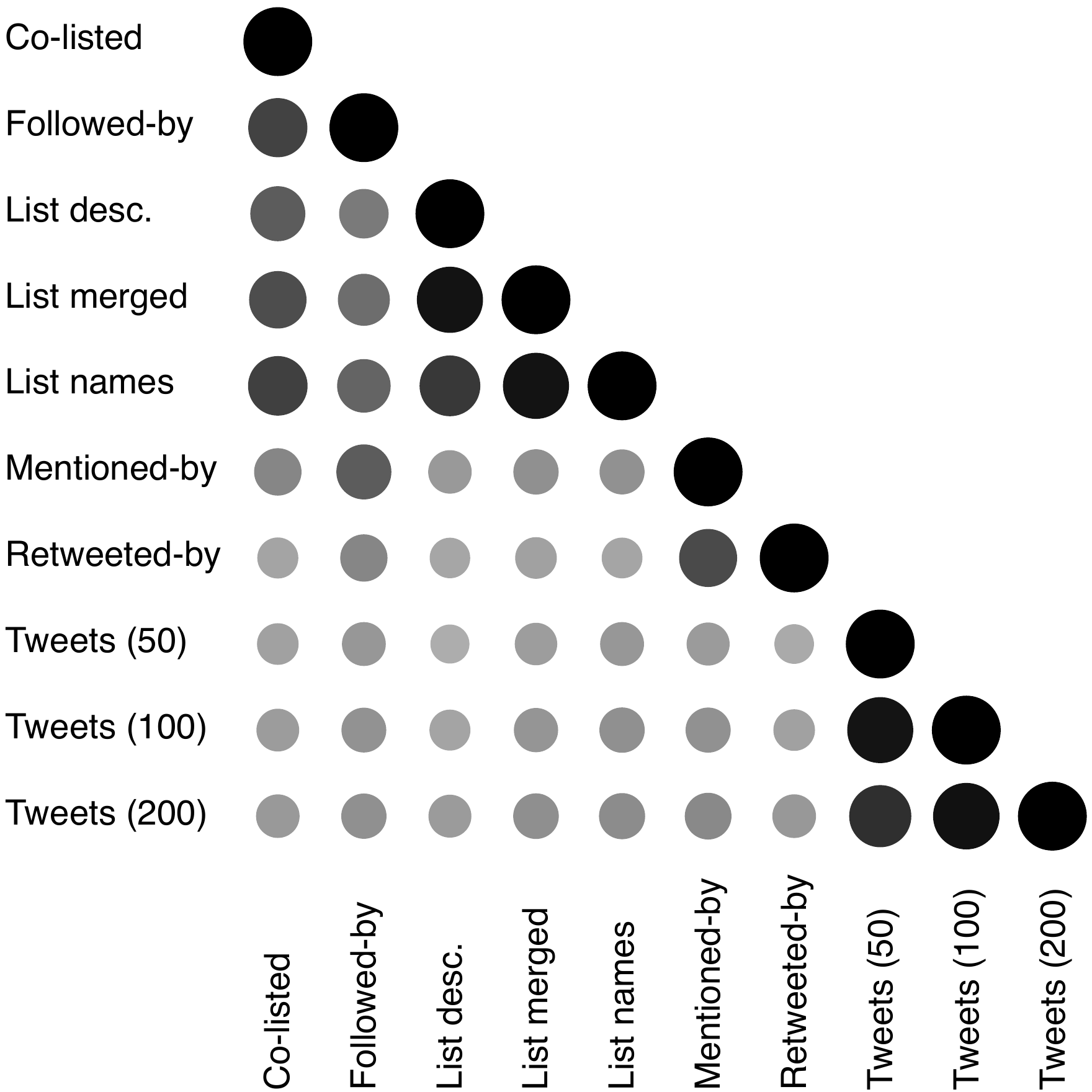}
\caption{Illustration of rank correlations between rankings generated using different criteria, averaged across all 10 datasets. A larger, more saturated point indicates a higher level of correlation.}
\label{fig:cor}
\end{figure}

Firstly, to examine the diversity of recommendations produced by the various criteria, \reffig{fig:cor} illustrates the agreement between rankings generated across all $10 \times 250$ runs, in terms of their pairwise Spearman rank correlations. These aggregated correlations indicate that there are a number of distinct signals present across different views of the same data. As we would expect, the different tweet profile sets are very highly-correlated. However, the rankings produced by list content text (\ie list names and descriptions) are considerably different, and correlated far more highly with the corresponding list memberships (\ie rankings generated on the co-listed graph). The latter criterion also shares some similarity with another network view, provided by the followed-by criterion.

\begin{table}[!t]
\centering
\caption{Total percentage of times that each criterion achieved first, second, and third place in terms of \emph{precision}, summed across all 10 datasets and each value of $k \in [10,50]$.}
\begin{tabular}{|l|ccc|}\hline
\bf Criterion & \bf First & \bf Second & \bf Third \\ \hline
Mentioned-by & 28\% & 8\% & 14\% \\ 
Followed-by & 24\% & 10\% & 12\% \\ 
Co-listed & 12\% & 18\% & 8\% \\ 
List names & 12\% & 14\% & 6\% \\ 
Tweets (200) & 10\% & 28\% & 6\% \\ 
List merged & 6\% & 10\% & 18\% \\ 
Tweets (100) & 4\% & 8\% & 12\% \\ 
List descriptions & 2\% & 4\% & 12\% \\ 
Tweets (50) & 2\% & 0\% & 12\% \\ 
Retweeted-by & 0\% & 0\% & 0\% \\ \hline
\end{tabular}
\label{tab:individual-prec}
\end{table}

\begin{table}[!t]
\centering
\caption{Total percentage of times that each criterion achieved first, second, and third place in terms of \emph{recall}, summed across all 10 datasets and each value of $k \in [10,50]$.}
\begin{tabular}{|l|ccc|}\hline
\bf Criterion & \bf First & \bf Second & \bf Third \\ \hline
Mentioned-by & 28\% & 8\% & 12\% \\ 
Followed-by & 24\% & 10\% & 12\% \\ 
Co-listed & 12\% & 18\% & 10\% \\ 
List names & 12\% & 14\% & 6\% \\ 
Tweets (200) & 10\% & 28\% & 6\% \\ 
List merged & 6\% & 10\% & 18\% \\ 
Tweets (100) & 4\% & 8\% & 12\% \\ 
List descriptions & 2\% & 4\% & 12\% \\ 
Tweets (50) & 2\% & 0\% & 12\% \\ 
Retweeted-by & 0\% & 0\% & 0\% \\ \hline
\end{tabular}
\label{tab:individual-rec}
\end{table}

To compare the accuracy of the criteria, we rank the performance of each criterion on each dataset for each value of $k$ in terms of both precision and recall measures. \reftab{tab:individual-prec} shows the total percentage of times that each criterion achieved first, second, and third place in terms of precision, while \reftab{tab:individual-rec} shows analogous results for recall. The criteria are ranked by the first, then the second, then the third column. Firstly, we observe that the precision and recall results are frequently low by the standards of most recommendation tasks. This reflects the difficulty of the task -- Twitter data is inherently noisy \cite{shamma09debates}, and for the purposes of curation, it may sometimes be the case that one user may be substituted for another in terms of the information that they provide. Nonetheless, in conjunction with a human curator, the ability to achieve up to $\approx 0.7$ recall on a manually-curated list does suggest that list curation can provide benefit in terms of supporting the work of online news outlets.

\begin{figure}[!t]
\centering
\includegraphics[width=0.82\linewidth]{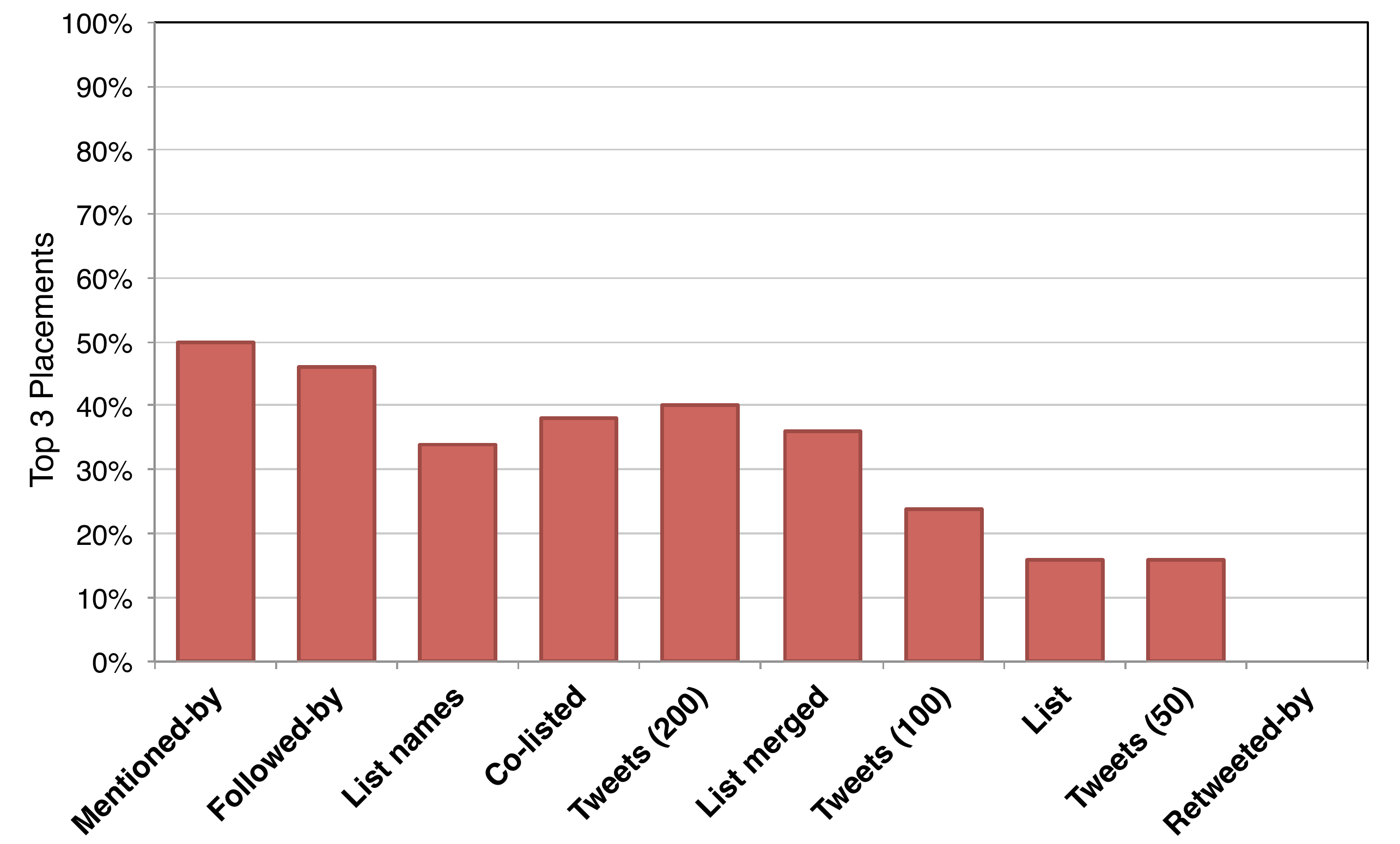}
\caption{Comparison of all individual criteria, for total percentage of top 3 placements, in terms of \emph{precision}, across all experiments.}
\label{fig:totalindividual}
\end{figure}

From the results, we see that the criteria derived from the analysis of the user mentions and followers networks are most successful, followed by the criteria derived from user list names and membership co-listings. The tweet content-based measures perform surprisingly poorly, although, as one might expect, the addition of more tweets does provide additional information and yield better results. Finally, the retweeted-by criterion does not achieve a top three placement in any of the 2,500 experiments. The sparsity of the retweet network, based on the number of retweets in the tweets collected for these datasets, appears to significantly limit the effectiveness of this criterion. Although $\approx20\%$ of all posts collected were retweets, many of these originated from users outside of the expanded datasets.

\begin{figure*}[!t]
\centering
\subfigure[Idaho]{
\includegraphics[width=0.435\linewidth]{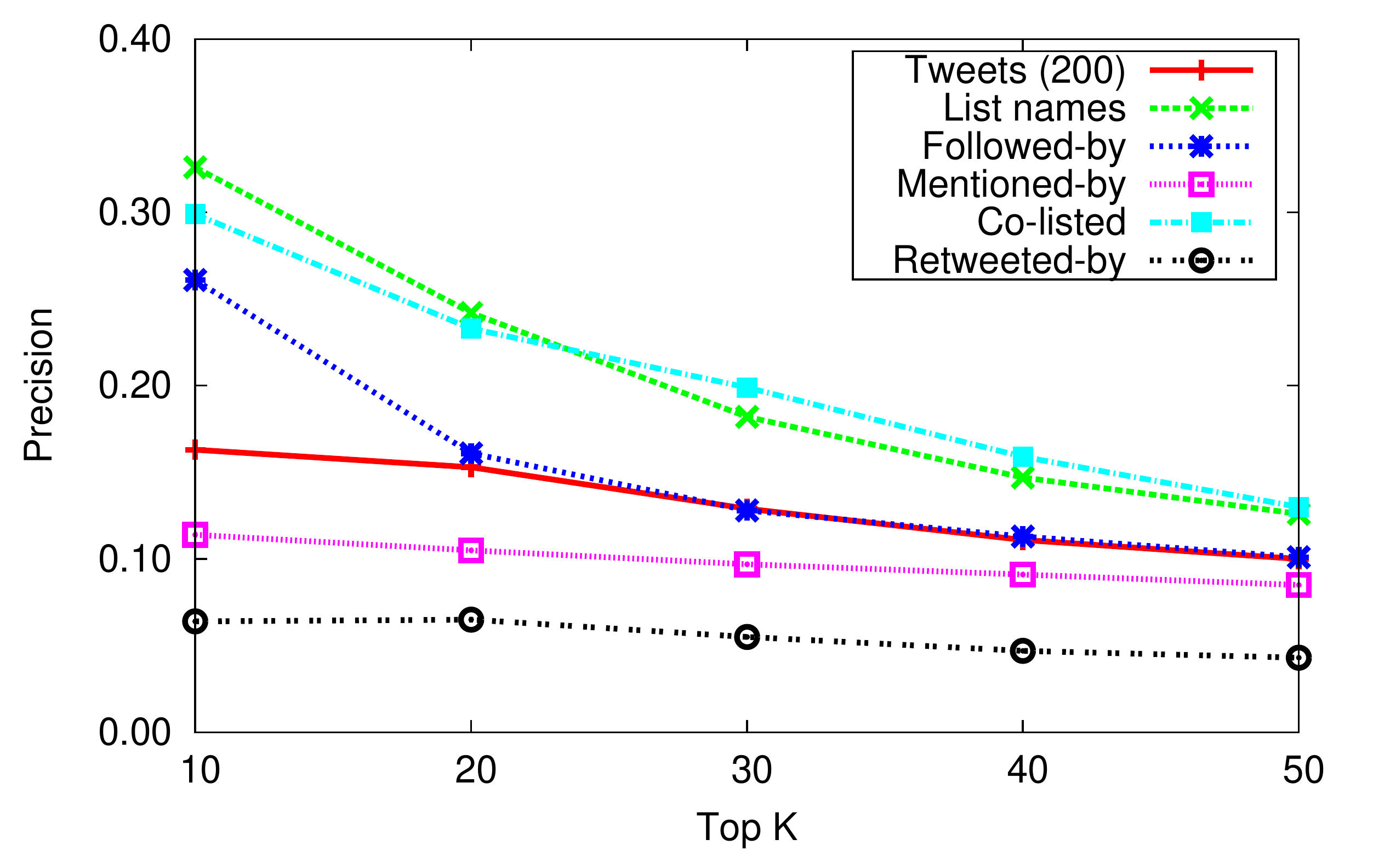}
}
\subfigure[Ohio]{
\includegraphics[width=0.435\linewidth]{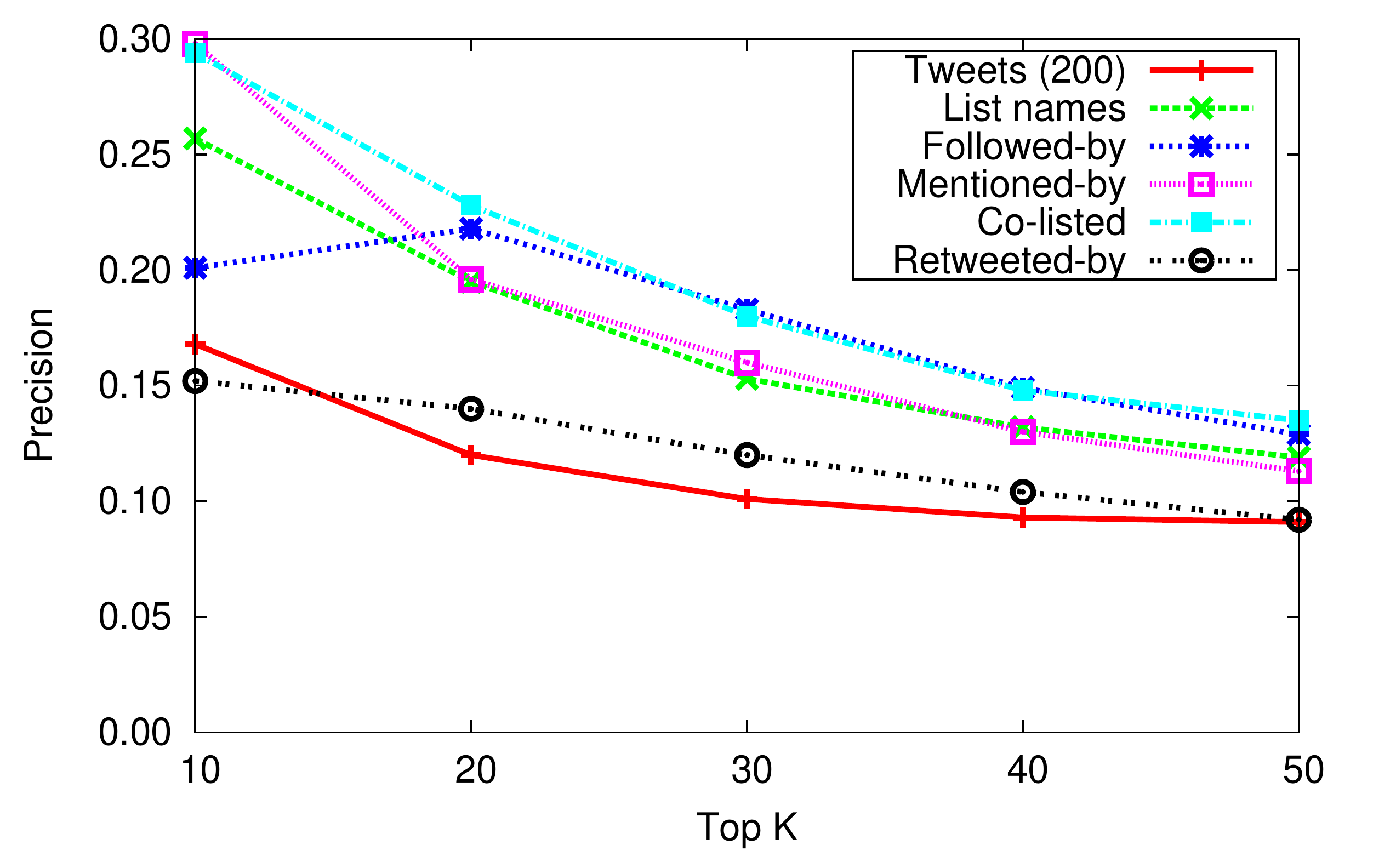}
}
\subfigure[North Dakota]{
\includegraphics[width=0.435\linewidth]{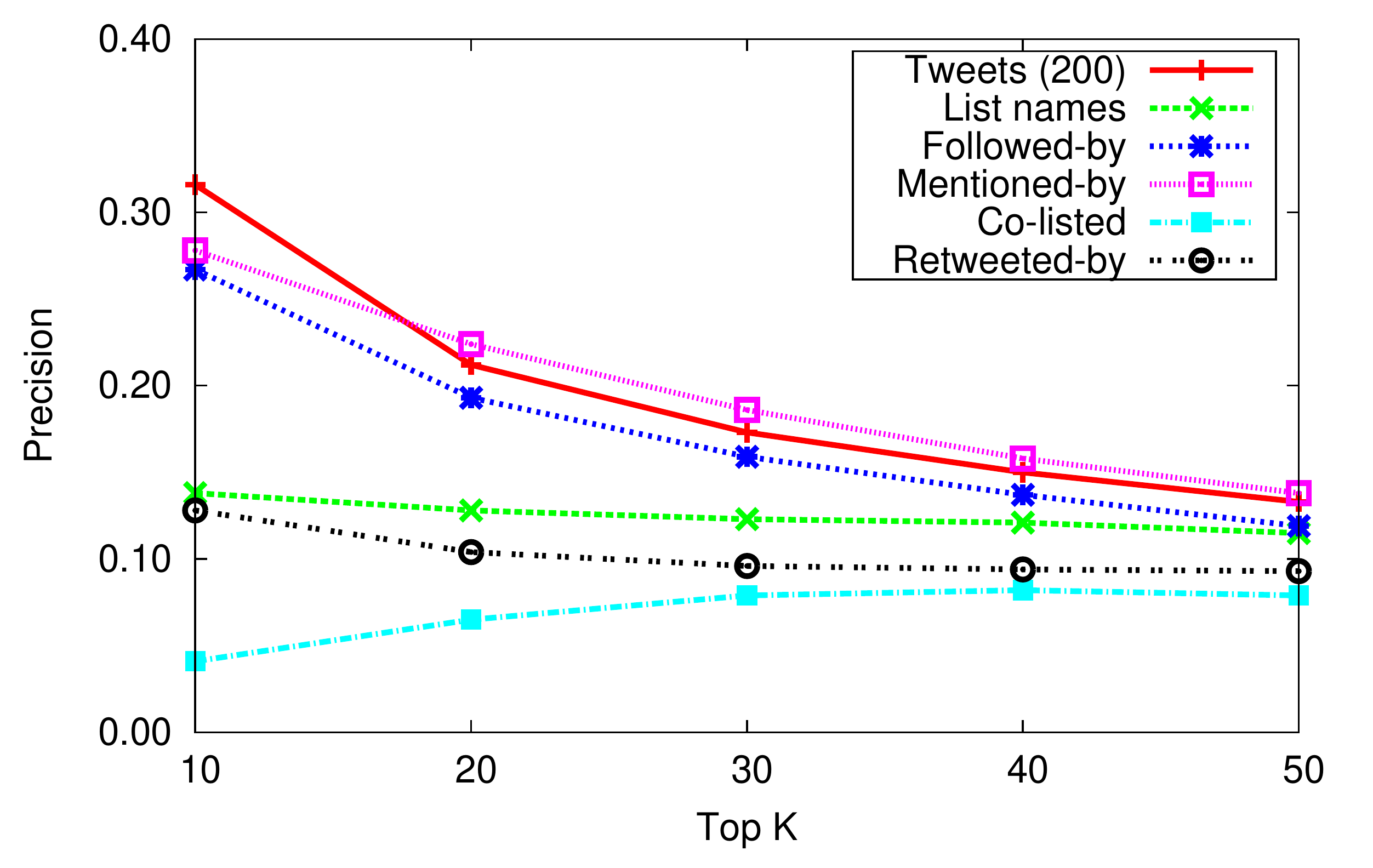}
}
\subfigure[Virginia]{
\includegraphics[width=0.435\linewidth]{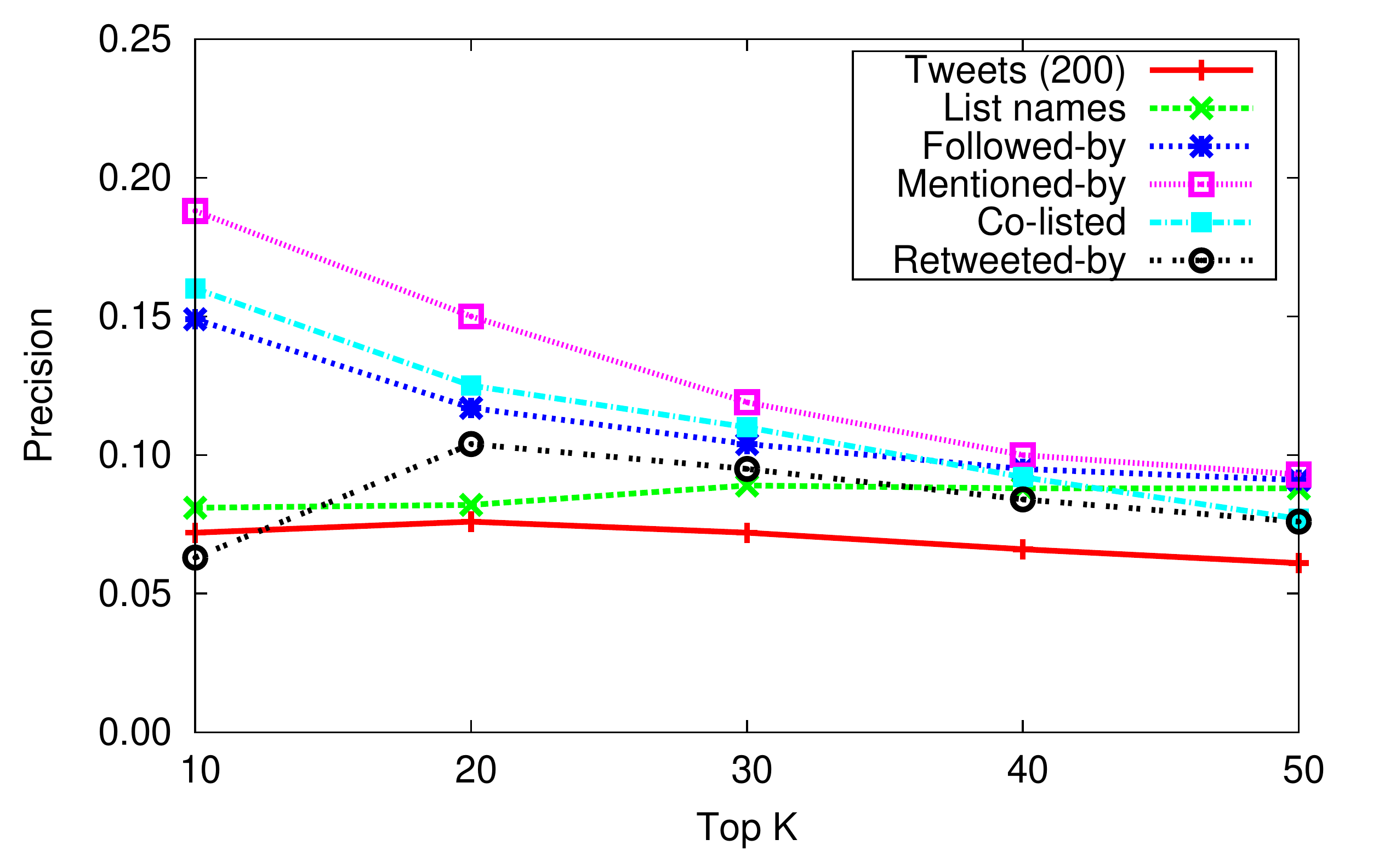}
}
\caption{Comparison of \emph{precision} scores for top $k \in [10,50]$ recommendations.}
\label{fig:precision}
\end{figure*}

\begin{figure*}[!t]
\centering
\subfigure[Idaho]{
\includegraphics[width=0.435\linewidth]{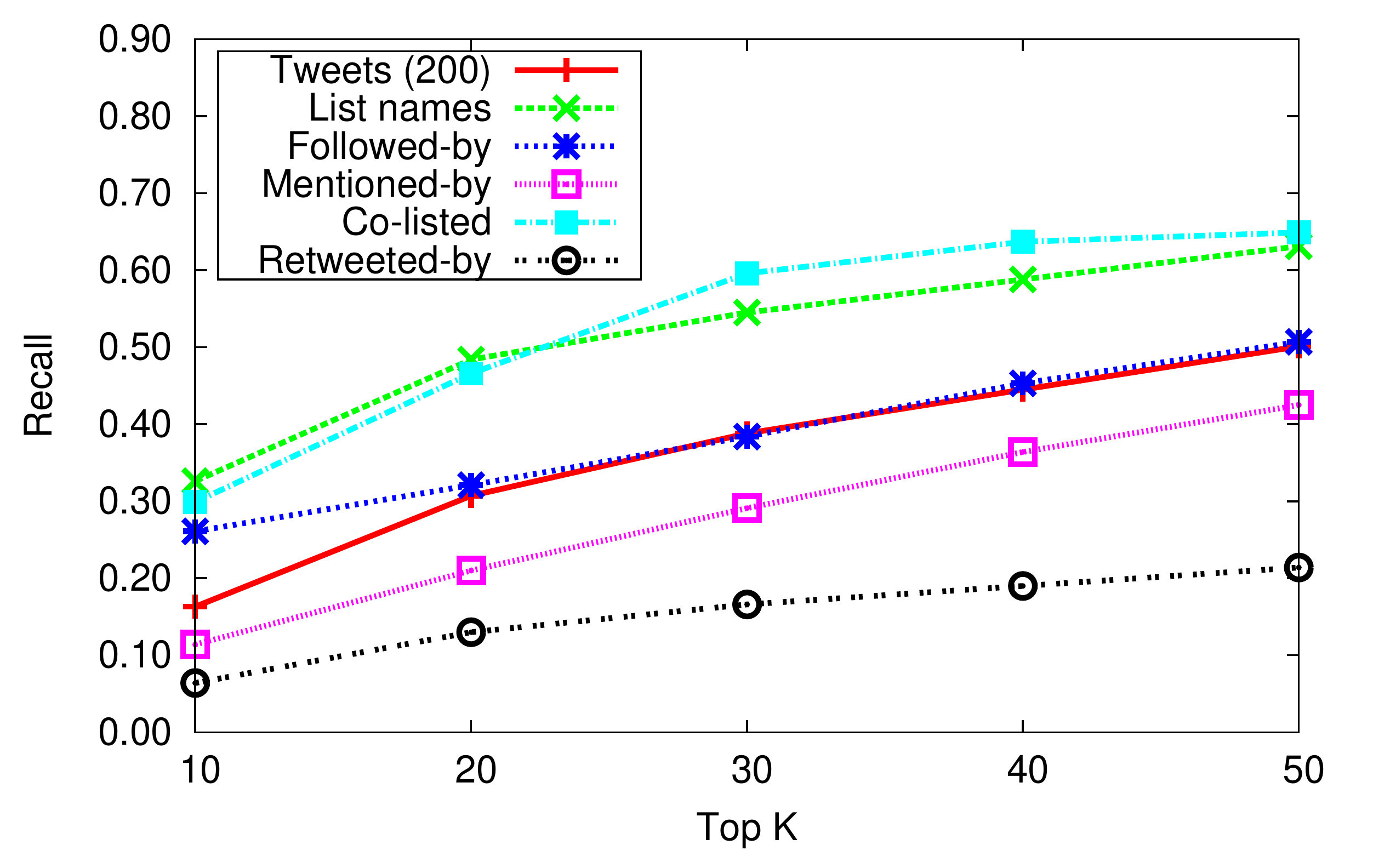}
}
\subfigure[Ohio]{
\includegraphics[width=0.435\linewidth]{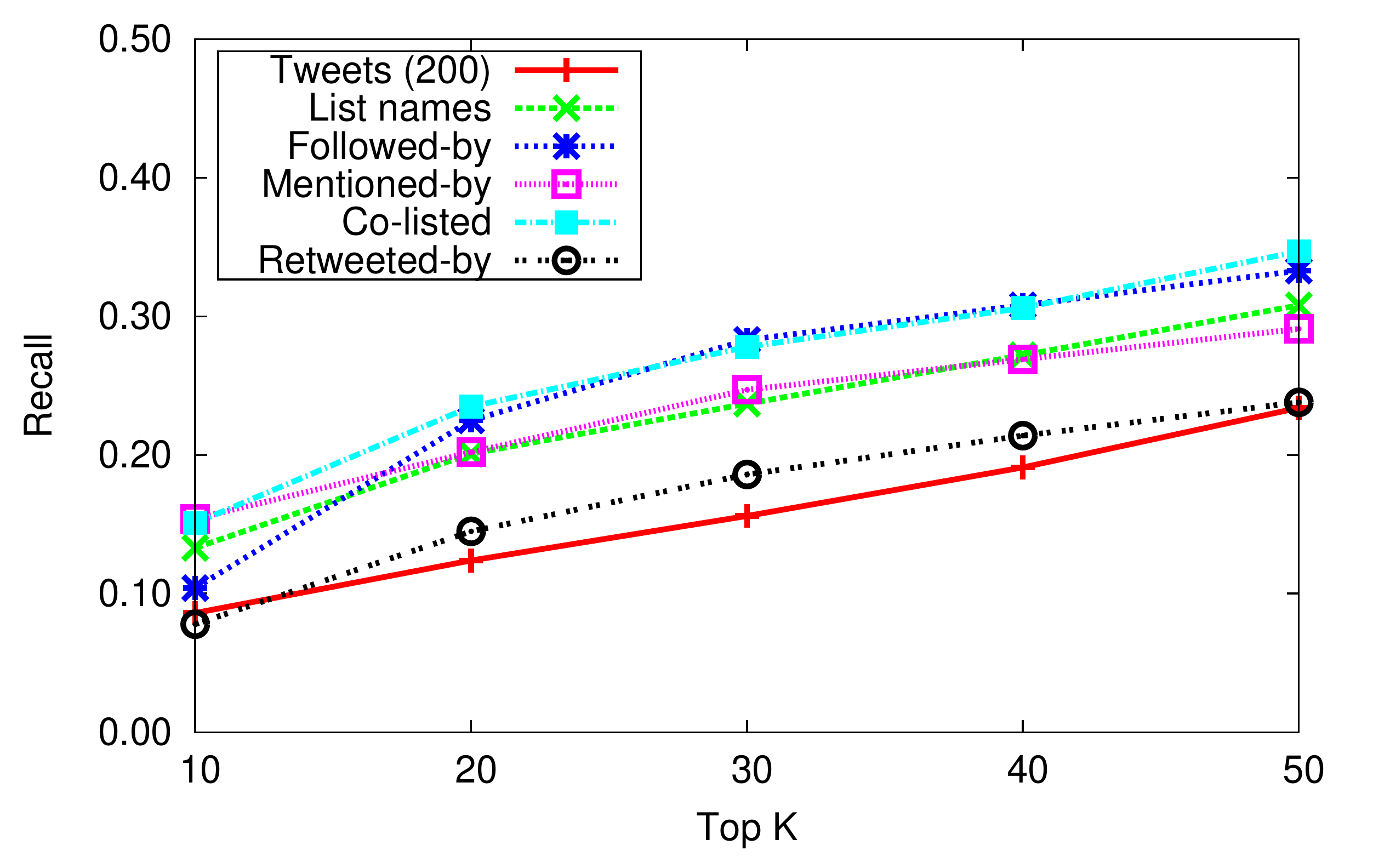}
}
\subfigure[North Dakota]{
\includegraphics[width=0.435\linewidth]{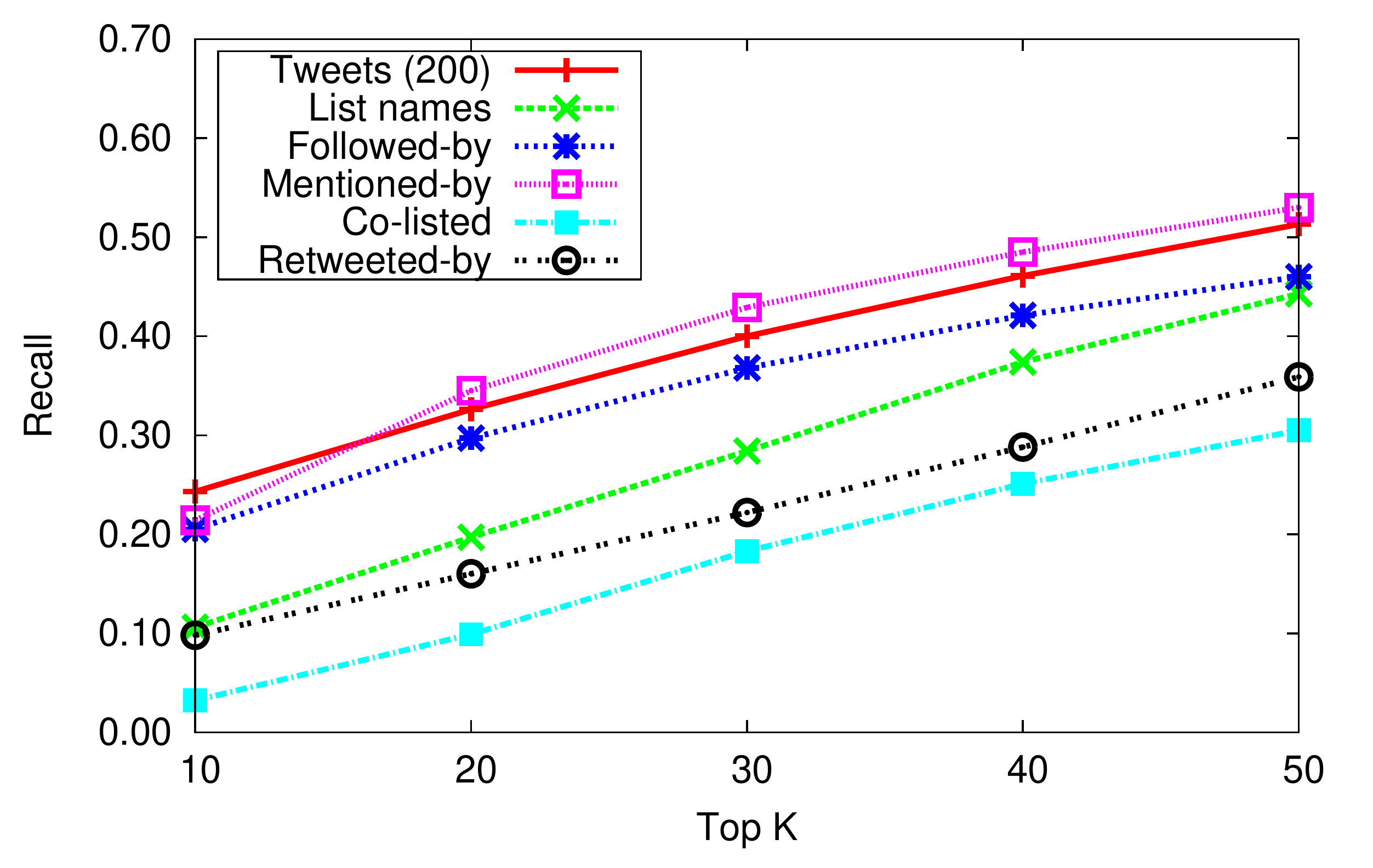}
}
\subfigure[Virginia]{
\includegraphics[width=0.435\linewidth]{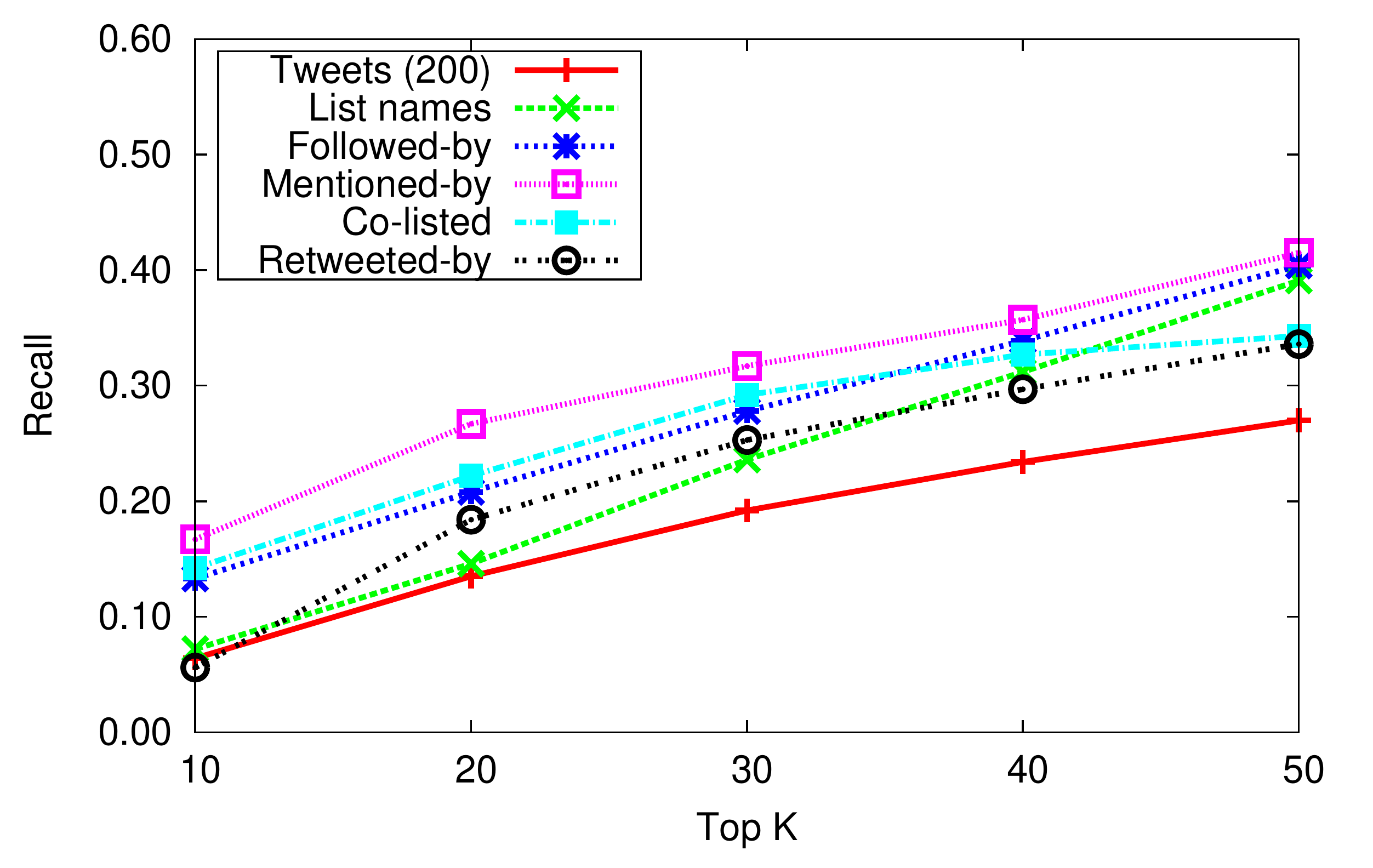}
}
\caption{Comparison of \emph{recall} scores for top $k \in [10,50]$ recommendations.}
\label{fig:recall}
\end{figure*}

\reffig{fig:totalindividual} shows the top percentage of times that each criterion achieved a top three placement across all experiments. It is clear that no individual criterion performs consistently-well across all ten datasets. In fact, the best performing criterion (mentioned-by) only achieves a top three placement $50\%$ of the time. This variation across datasets suggests that no single criterion or view alone is sufficient to support list curation. For instance, the tweet profile approach is successful on the Alaska and Georgia datasets, yet is the least accurate criterion on a number datasets (\eg Ohio, Virginia -- see Figures \ref{fig:precision} and \ref{fig:recall}). Similarly, co-listed information proves informative on the Idaho and Ohio datasets, yet achieves a precision of $< 0.10 \; \forall k$ in the case of North Dakota.


\begin{figure}[!b]
\centering
\includegraphics[width=0.85\linewidth]{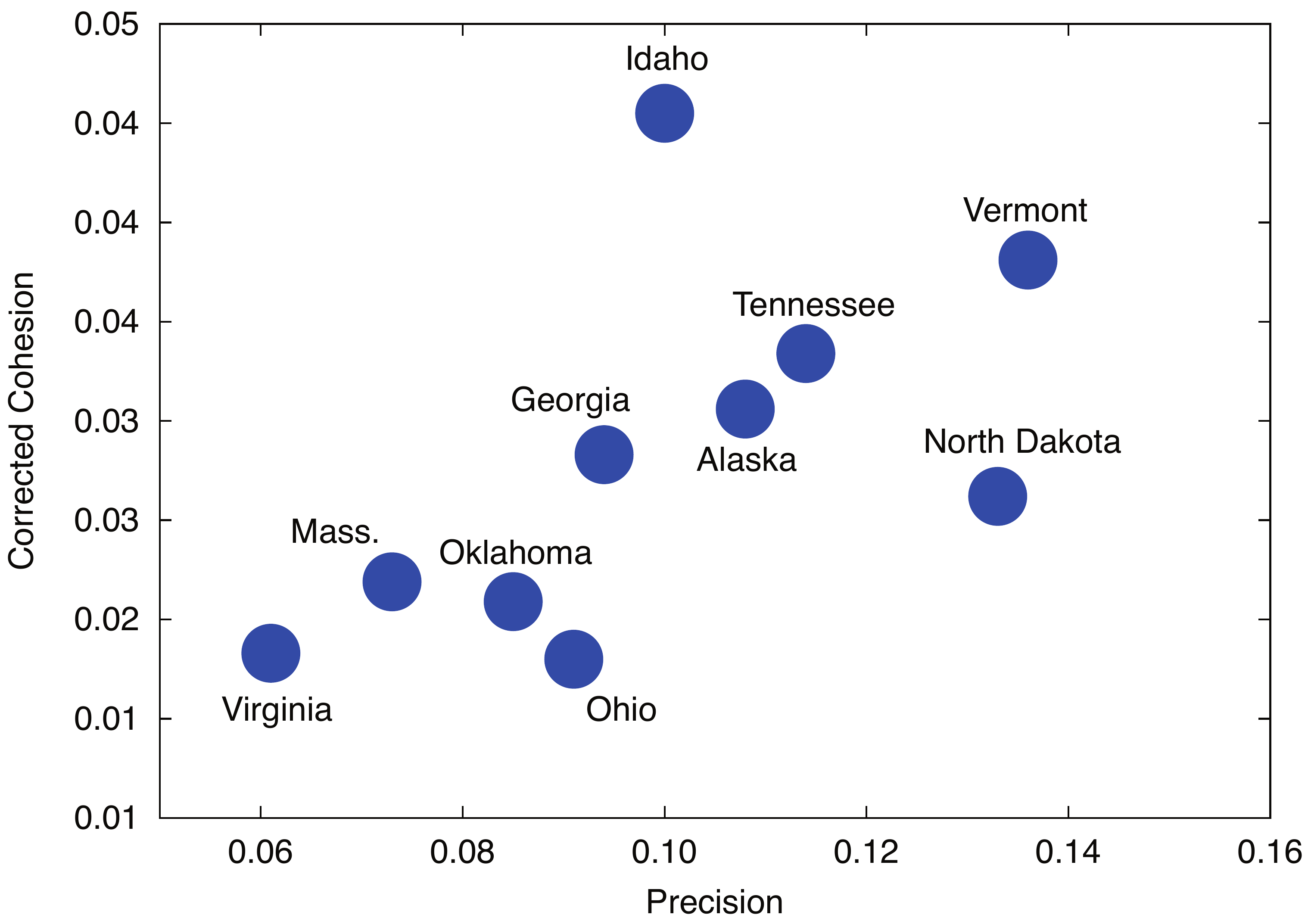}
\caption{Plot of corrected seed \emph{cohesion} versus \emph{precision} for ten datasets for top $k=50$ recommendations based on the {\emph{tweet profile (200)}} criterion.}
\label{fig:cohtweets}
\end{figure}

\begin{figure}[!t]
\centering
\includegraphics[width=0.85\linewidth]{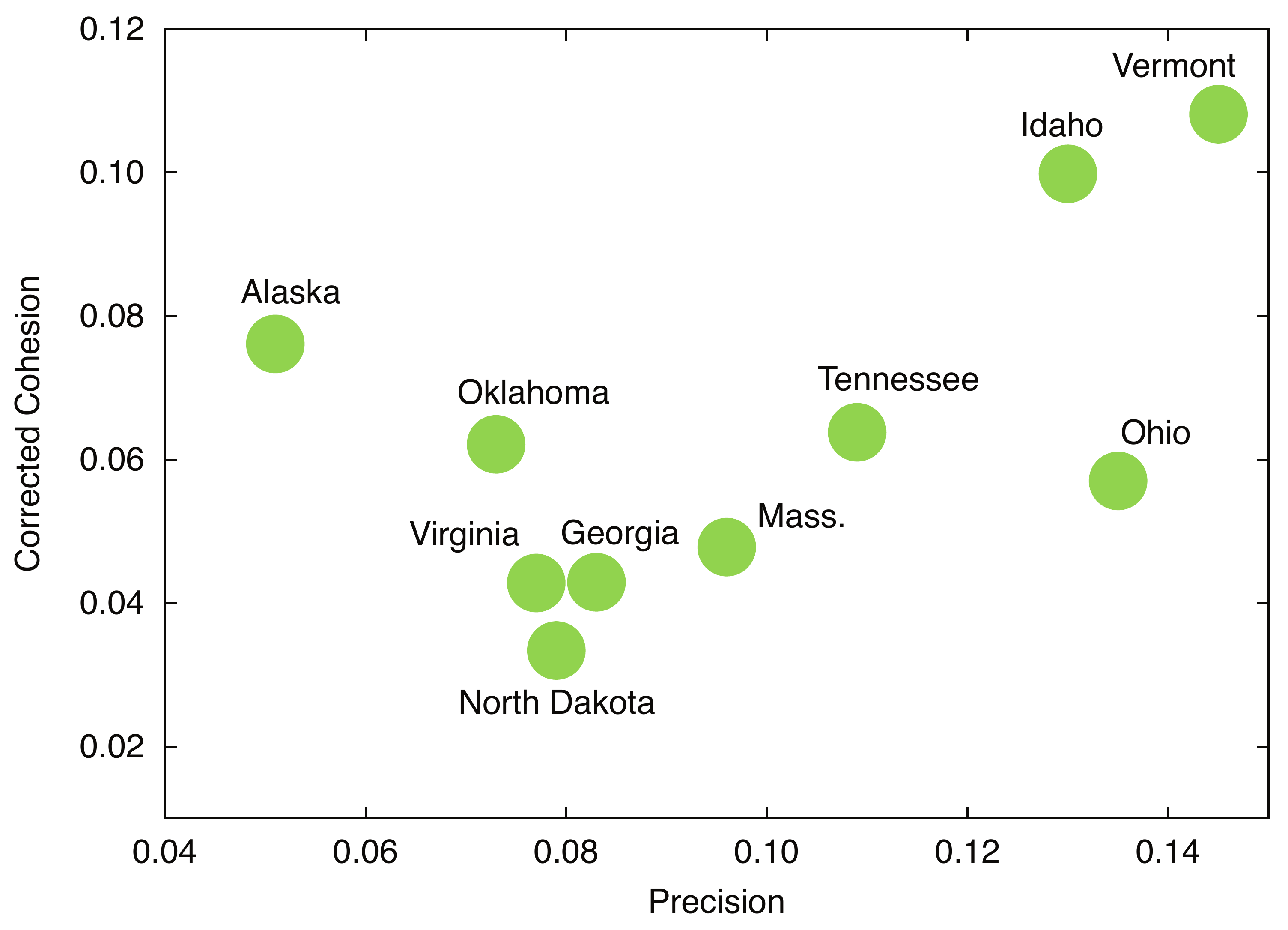}
\caption{Plot of corrected seed \emph{cohesion} versus \emph{precision} for ten datasets for top $k=50$ recommendations based on the {\emph{co-listed}} criterion.}
\label{fig:cohcolist}
\end{figure}

To investigate the quality of information provided by individual views, relative to the annotated seed set, we consider the \emph{cohesion} of that set of users for each criterion as follows. For a given criterion, we  compute the mean pairwise similarity between users in the seed set -- since we make use of cosine similarity in all cases, this value has the range $\in [0,1]$. We then compute the mean expected similarity for a set of users of that size as follows: we re-label the user identifiers in the full dataset, and compute the mean pairwise similarity the new seed set users. This process is repeated over a large number of randomised runs, yielding an approximation of the expected value. We then employ the widely-used adjustment technique introduced by Hubert \& Arabie \cite{hubert85compare} to correct for chance agreement: 
\[
\textrm{CorrectedCohesion} = \frac{\textrm{Cohesion} - \textrm{ExpectedCohesion}}{1 - \textrm{ExpectedCohesion}}
\]
Figures \ref{fig:cohtweets} and \ref{fig:cohcolist} respectively show a plot of corrected cohesion, as calculated above, against precision for the top $k=50$ recommendations on all datasets. As one might expect, we see a strong correlation between the cohesiveness of the seed set in a given view, and the quality of recommendations produced on that view. 
For instance, in the case of Virginia in \reffig{fig:cohtweets}, we see that the mean similarity between seed users in terms of their Tweet profiles is little different than if we had selected a pair of users at random from the overall datasets -- naturally, our ability to identify relevant users based on their Tweet profiles alone is strictly limited here. Similarly, in \reffig{fig:cohcolist} we see that list memberships do not effectively distinguish seed from non-seed users in the case of North Dakota. Similar trends are evident in the case of the other criteria, for both precision and recall.

\subsection{Aggregating Multiple Criteria}
\label{sec:eval2}

As discussed previously, the various criteria presented here can potentially produce rankings of users that can differ considerably, and the effectiveness of these criteria can vary significantly from one Twitter dataset to another. To resolve this issue and to harness the diversity of views of Twitter views available (as indicated by \reffig{fig:cor}), we suggest the combination of rankings generated using both network and content-based techniques. To actually combine the rankings, we use SVD-based aggregation, which has previously been shown to be effective for combining recommendations in other contexts \cite{wu10recsys}. Specifically, we aggregate the top five performing criteria from \refsec{sec:eval2}. It is clear from the pairwise correlations in \reffig{fig:cor} that these represent a diverse set of network and content-based criteria:
\begin{itemize}
\item \emph{Network criteria:} Followed-by, mentioned-by, co-listed.
\item \emph{Content criteria:} Tweets (200), list names.
\end{itemize}

To compare the performance of SVD aggregation relative to the individual criteria, we repeat the cross-validation experiments, comparing SVD against the top five individual criteria used in the aggregation process. We then re-rank these alternative approaches based on their precision and recall, again computed as described in \refsec{sec:setup}.

\begin{table}[!t]
\centering
\caption{Comparison of performance of SVD aggregation versus top five individual criteria, in terms of top 3 \emph{precision} placements, across all experiments.}
\begin{tabular}{|l|ccc|}\hline
\bf Criterion & \bf First & \bf Second & \bf Third \\ \hline
SVD & 48\% & 28\% & 16\% \\ 
Followed-by & 20\% & 8\% & 14\% \\ 
Tweets (200) & 12\% & 2\% & 26\% \\ 
Co-listed & 10\% & 12\% & 20\% \\ 
List names & 6\% & 22\% & 6\% \\ 
Mentioned-by & 4\% & 28\% & 18\% \\ \hline
\end{tabular}
\label{tab:medals-prec}
\end{table}
\begin{table}[!t]
\centering
\caption{Comparison of performance of SVD aggregation versus top five individual criteria, in terms of top 3 \emph{recall} placements, across all experiments.}
\begin{tabular}{|l|ccc|}\hline
\bf Criterion & \bf First & \bf Second & \bf Third \\ \hline
SVD & 46\% & 30\% & 16\% \\ 
Followed-by & 22\% & 6\% & 14\% \\ 
Tweets (200) & 12\% & 2\% & 26\% \\ 
Co-listed & 10\% & 12\% & 22\% \\ 
List names & 6\% & 22\% & 4\% \\ 
Mentioned-by & 4\% & 28\% & 18\% \\ \hline
\end{tabular}
\label{tab:medals-rec}
\end{table}

\begin{figure}[!t]
\centering
\includegraphics[width=0.85\linewidth]{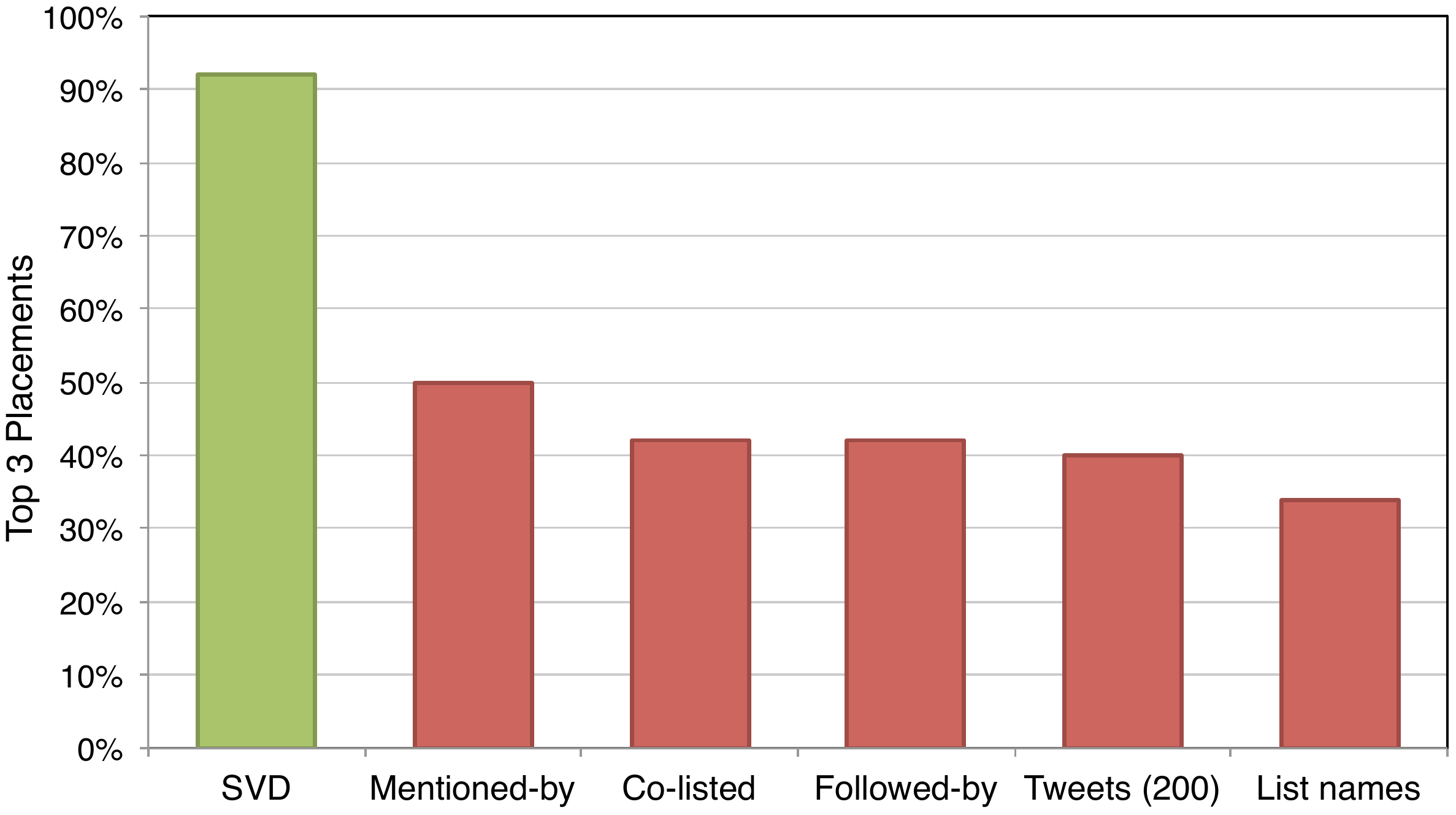}
\caption{Comparison of SVD aggregation versus individual criteria, for total percentage of top 3 placements, in terms of \emph{precision}, across all experiments.}
\label{fig:totalsvd}
\end{figure}

\reftab{tab:medals-prec} shows the the percentage of times that each approach achieved first, second, and third place, in terms of precision, as computed across all 10 datasets and each of the five values of $k$ that we examined (ranked by first column, then second, then third). We observed that SVD-based aggregation consistently out-performed the other techniques, achieving first place in almost half of the experiments, and finishing in the top three in $92\%$ of the experiments. In contrast, from \reffig{fig:totalsvd} we see that the most competitive individual criterion on this ordering (mentioned-by) finished in the top three during only $50\%$ of the experiments. Again content-based techniques fare relatively poorly. We see similar performance in terms of recall, as evidenced by the ordering of approaches in \reftab{tab:medals-rec}.


\section{Conclusions}
The problem of content curation in social media networks is becoming increasingly important, particularly in the context of news curation for media outlets. In the case of Twitter, curating a list of authoritative users tweeting about a given news story provides a means of monitoring discussions around that story. However, currently this is a time-consuming manual task. Here we presented a range of criteria for building topical user lists, based on an initial seed set. By analysing the cohesion of the training data across different views of the same datasets, we demonstrated the strengths and weaknesses of these recommendation criteria, in the context of a limited availability of Twitter data. To overcome the weaknesses, we proposed the use of SVD rank aggregation. Experiments on a range of Twitter datasets relating to US politics demonstrated that this aggregation process yields more robust recommendations, succeeding in cases where individual content- or network-based criteria perform poorly.


\vspace{3 mm}\noindent\emph{Acknowledgments.} 
This research was supported by Science Foundation Ireland Grant 08/SRC/I1407 (Clique: Graph and Network Analysis Cluster). The authors also thank Storyful for their participation in the evaluation.

\bibliographystyle{abbrv}
\bibliography{curation} 

\end{document}